\documentclass[pss]{wiley2sp}
\usepackage{amsmath}
\usepackage{bm}              
\usepackage{w-greek}         

\begin{document}


\title{On the concept of effective temperature in  current carrying quantum
critical states}

\titlerunning{Effective temperatures in current carrying states}

\author{%
  Stefan Kirchner \textsuperscript{\Ast,\textsf{\bfseries 1}},
  Qimiao Si \textsuperscript{\textsf{\bfseries 2}}}
\authorrunning{S.~Kirchner et al.}

\mail{e-mail
  \textsf{kirchner@pks.mpg.de}, Phone:
  +49-351-871-1121}

\institute{%
  \textsuperscript{1}\,Max Planck Institute for the Physics of Complex Systems, 
N\"{o}thnitzer Str. 38, 
D-01187 Dresden, Germany\\
  \textsuperscript{2}\,Department of Physics \& Astronomy, Rice University, Houston, TX, 77005, USA\\
}

\received{XXXX, revised XXXX, accepted XXXX} 
\published{XXXX} 


\abstract{%
%
%
%
\abstcol{Quantum criticality 
has attracted considerable attention  both theoretically 
and experimentally as a way to describe 
part of the phase diagram of strongly
correlated systems. A scale-invariant fluctuation spectrum at a quantum 
critical point (QCP) implies the absence of any 
intrinsic scale. Any experimental probe may therefore create 
an out-of-equilibrium setting; the system would be in a non-linear 
response regime, which violates the fluctuation-dissipation 
theorem (FDT). 
}
{
Here, we study this violation and related out-of 
equilibrium phenomena in a single electron transistor (SET) with 
ferromagnetic leads, which can be tuned through a quantum phase
transition. We review the breakdown of the FDT and study the universal
behavior of the fluctuation dissipation relation of various correlators
in the quantum critical regime. In particular, we
explore the concept of effective temperature as a means to extend 
the FDT into the non-linear
regime.
}}

\maketitle   

\section{Introduction}
Quantum criticality has become a new paradigm in describing complex quantum matter.
It occurs when matter goes continuously from one phase  to another  at zero temperature
as a function of a control parameter. No classification scheme 
into universality classes has  emerged so far,
but it has become clear that not all QCPs
are characterized by order parameter fluctuations alone. 
Rather, inherently quantum modes
may become critical and must also be incorporated in the description.
Such a QCP 
is not described by a 
Ginzburg-Landau-Wilson functional
and has 
no classical (finite temperature)
counterpart. The strongest evidence to date for the existence of such unconventional quantum criticality  
has come from heavy fermion systems at the border to magnetism. 
One of the potential
mechanisms for quantum-critical
heavy fermion metals, that is beyond the Ginzburg-Landau-Wilson paradigm,
is the  critical destruction of the Kondo 
effect~\cite{Coleman.01,Si.01,Loehneysen.07,Gegenwart.08}.
This destruction is local in space and leads to interacting critical modes 
along imaginary time.\\ 
Out-of-equilibrium states near quantum phase transitions, however, have so far received only limited
theoretical attention despite a long-standing strong interest in their classical
counterparts. 
Experimentally, on the other hand, it seems unavoidable to generate out-of-equilibrium states during
a measurement in the quantum critical regime. This regime is characterized by the absence of any 
intrinsic scale.  Any measurement may therefore potentially 
perturb the system beyond the linear response regime where the 
FDT no longer holds\footnote{The FDT states that, for a small enough
external drive, the system will react in a way prescribed 
by its equilibrium fluctuations.}.
Recently, we demonstrated
that a local quantum critical point can be realized in a SET
attached to ferromagnetic leads~\cite{Kirchner.05a}:
\begin{figure*}[ht!]%
  \sidecaption
  \includegraphics*[width=.68\textwidth,height=4cm]{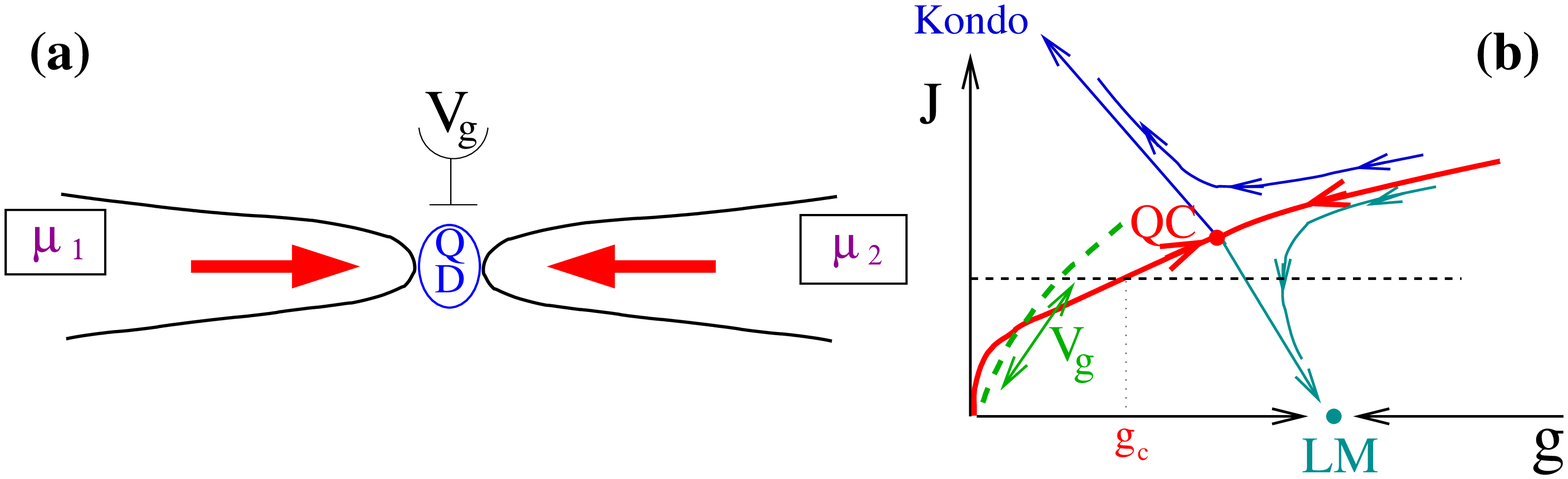}%
  \caption[]{%
(a) Principal setup of the magnetic SET. The arrows in the left and right 
leads indicate the magnetization 
in the leads. For antiparallel alignment, $h_{\mbox{\tiny loc}}$ in Eq.(\ref{hamiltonian-bfk-n=2}) vanishes.
(b) Phase diagram: the Fermi liquid phase ('Kondo') and the critical local moment ('LM') phase are
separated by a QCP. 
The gate voltage $V_G$ allows to tune the system through a quantum phase transition.}
    \label{sidecaption}
\end{figure*}
As the  applied gate voltage is varied, 
the magnetic SET can be tuned through a
continuous quantum phase transition
where the Kondo effect is critically destroyed~\cite{Kirchner.05a,Kirchner.08b}.
Such a nano
device also constitutes a simplified system, both
theoretically and experimentally, to study well-defined out-of-equilibrium
states that give rise to unique steady-state limits. 
In order to elaborate on this insight, we identified a limit
where both in- and out-of equilibrium properties 
of the magnetic SET can be determined exactly~\cite{Kirchner.09}. 
We studied the non-linear current-voltage (or I-V) characteristics 
for current-carrying steady states across the quantum transition~\cite{Kirchner.09}.\\
In the present work, we elaborate on the fate of the fluctuation-dissipation relation in the 
non-linear regime
and explore the notion of effective temperature.


\section{The quantum critical  single-electron transistor}
The magnetic SET, a quantum dot attached to ferromagnetic leads, has been investigated experimentally 
and theoretically~\cite{Pasupathy.04,Martinek.03,Choi.04,Kirchner.05a,Kirchner.08b}. 
The couplings of the local degrees of freedom to the conduction electrons and
to the magnons in the leads allow for a dynamical competition between the Kondo singlet formation
and the magnon drag. As a result,
the low-energy properties are governed by a 
Bose-Fermi Kondo model (BFKM)~\cite{Kirchner.05a,Kirchner.08b}.
\begin{eqnarray}
{\mathcal H}_{\mbox{bfk}}&=&
\sum_{\stackrel{i,j}{k,k^\prime,\sigma,\sigma^\prime}}
J_{i,j}{\bf S} \cdot
c^{\dagger}_{k,\sigma,i}\frac{{\bf \sigma}}{2}c^{}_{\sigma^\prime,j} \\
&+&
\sum_{{\bf k},i,\sigma} 
\tilde{\epsilon}_{{\bf k}\sigma i}^{}~
c_{{\bf k}\sigma i}^{\dagger} c_{{\bf k} \sigma i} + 
h_{\mbox{\tiny loc}}
S_{z} 
\\
&+&
g 
\sum_{\beta,{\bf q},i} S_{\beta} 
(\phi_{\beta,{\bf q},i} +
\phi^{\dagger}_{\beta,{\bf q},i} )
+ \sum_{\beta,{\bf q},i}
\omega_{\bf q}^{}\,
\phi_{\beta,{\bf q},i}^{\;\dagger} \phi_{\beta,{\bf q},i}.\nonumber
\label{hamiltonian-bfk-n=2}
\end{eqnarray}
where $i,j\varepsilon \{L,R\}$
and  
$h_{\mbox{\tiny loc}}
= g \sum_i m_i$, is a local magnetic field  with
$m_L/m_R$ being the ordered moment of the left/right
leads. For antiparallel alignment and equal couplings one finds  $m_L=-m_R$.
$\tilde{\epsilon}_{{\bf k}\sigma i}$ is the 
Zeeman-shifted conduction electron dispersion,
and ${\phi}_{\beta,i}$, with $\beta = x,y$, describes
the magnon excitations.
The spectrum of the bosonic modes is determined by the density of states of the magnons,
$\sum_q \left [
\delta(\omega-\omega_q)-\delta(\omega+\omega_q)
\right ] \sim |\omega|^{1/2}
sgn(\omega)$ up to some cutoff $\Lambda$. 
A sketch of the magnetic SET is shown in Fig.~\ref{sidecaption}(a). The resulting
phase diagram of the system is displayed in 
Fig.~\ref{sidecaption}(b). For further details, see~\cite{Kirchner.05a,Kirchner.08b,Kirchner.09}.
Nonequilibrium states are created by  keeping 
left and right lead at different chemical potentials but the same temperature $T$,
($eV=\mu_L-\mu_R$).
The current between the dot and the right lead (R) for an arbitrary bias voltage 
$V$ is \cite{Meir.92} 
\begin{eqnarray*}
I_R\!\!&=\!\!&\frac{i e}{\hbar}\!\int\!\! d\omega \rho_R(\omega)
\big 
[f_R(\omega)({\mathcal{T}}^r_{RR}(\omega)-{\mathcal{T}}^a_{RR}(\omega))
+{\mathcal{T}}^<_{RR}(\omega) \big ],
\label{eq:current}
\end{eqnarray*}
where ${\mathcal{T}}_{\alpha,\beta},~(\alpha,\beta=L/R)$ 
is the T-matrix of the BFKM, $f_{L/R}(\omega)=f(\omega_{ }\pm \mu_{L/R})$ is the
Fermi function for the left/right lead and
${\mathcal{T}}^{r/a/<}$ denote the retarded, advanced and lesser T-matrix respectively.
These three functions together define the ``larger'' function:
${\mathcal{T}}^{>}={\mathcal{T}}^{r}-{\mathcal{T}}^{a} +{\mathcal{T}}^{<}$
and the corresponding fluctuation-dissipation ratio (FDR) becomes
\begin{equation}
FDR_{\mathcal{T}} \equiv \frac{{\mathcal{T}}^>+{\mathcal{T}}^<}{{\mathcal{T}}^>-{\mathcal{T}}^<},
\label{FDR_definition}
\end{equation} 
which, in equilibrium, reduces to 
$1-2f(\omega)=\tanh(\omega/2T)$ (or to $1+2b(\omega)=\coth(\omega/2T)$,
in the case of bosonic variables), in accordance with the FDT.
Eq.(\ref{FDR_definition}) implies
\begin{equation}
{\mathcal{T}}^{<}=\frac{1}{2}(FDR_{\mathcal{T}}-1)[{\mathcal{T}}^{r}-{\mathcal{T}}^{a}].
\end{equation}
The spin exchange matrix between the leads and the local moment arises out of virtual charge fluctuations.
We therefore have $J_{RR}=\alpha J_{LL}$, which implies ${\mathcal{T}}_{RR}=\alpha {\mathcal{T}}_{LL}$.
An equation  similar to the one for $I_R$  holds for the current between the left lead (L) and
the dot, $I_L$.
In the steady state limit, these two have to be equal: 
$I_L+I_R=0$. (We will only consider the steady state limit here,
where correlation functions  depend on time differences: $<A(t_j)B(t_i)>=<A(t_j-t_i)B(0)>$.)
For simplicity, we take $\rho_R(\omega)=\rho_L(\omega)=\rho(\omega)$.
The steady state condition then implies 
\begin{eqnarray*}
&&\int\!\! d\omega \rho(\omega)
\big 
[f_L(\omega)+\alpha f_R(\omega)+\frac{1+\alpha}{2}(FDR_{\mathcal{T}}-1)
\big ] \\
&&\times ({\mathcal{T}}^r_{LL}(\omega)-{\mathcal{T}}^a_{LL}(\omega))=0,
\end{eqnarray*}
and since $\rho(\omega)\geq 0$, 
${\mathcal{T}}^r_{LL}(\omega)-{\mathcal{T}}^a_{LL}(\omega)=2 \mbox{Im}{\mathcal{T}}(\omega) \geq 0$,
once can solve for $FDR_{\mathcal{T}}$:
\begin{eqnarray}
FDR_{\mathcal{T}}(\omega,T,V) = 
\frac{\sinh(\omega/T)}{
\cosh(\omega/T)+\cosh(V/2T)
},
\label{fdr_g}
\end{eqnarray}
where we assumed equal couplings ($\alpha=1$).
This equation was used in~\cite{Kirchner.09} to infer the scaling properties of the T-matrix in
the quantum critical regime from the scaling of the $I-V$ characteristics.
\begin{figure*}[ht!]%
  \sidecaption
  \includegraphics*[width=0.68\textwidth,height=4cm]{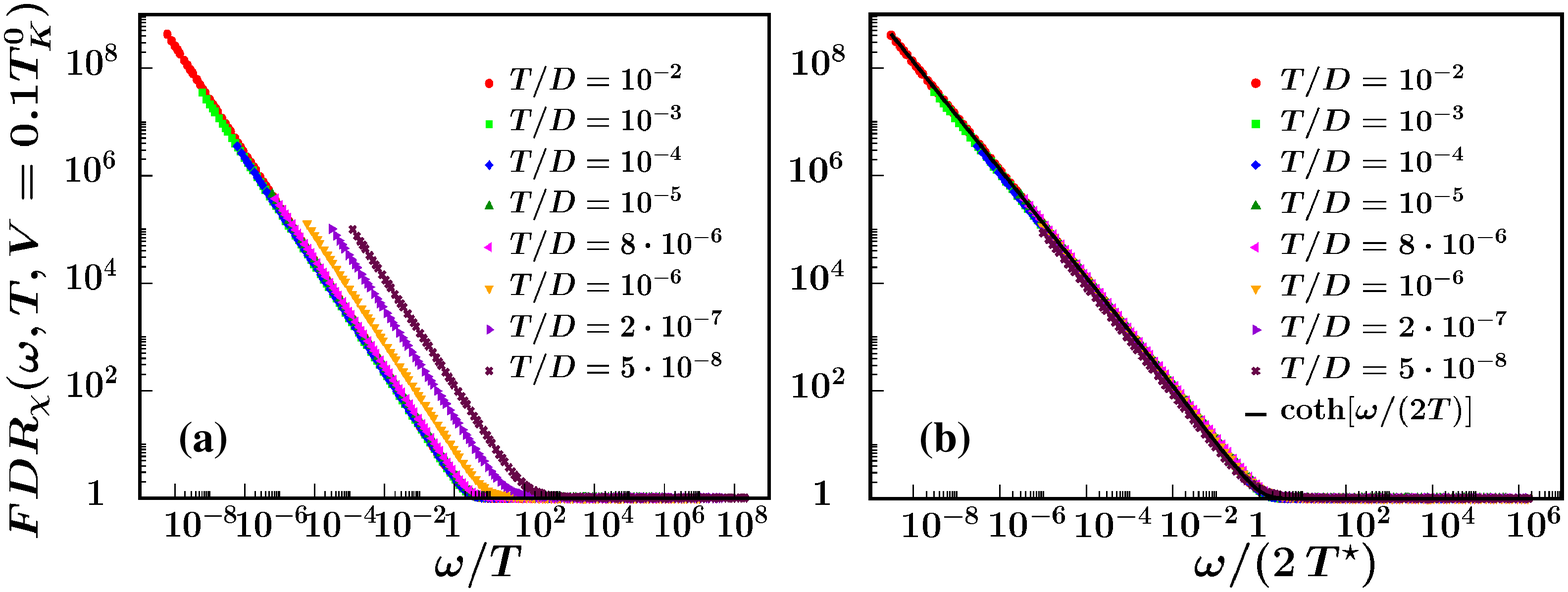}%
  \caption[]{(a) The fluctuation-dissipation ratio for the spin susceptibility in the
critical local moment phase ($g=4\cdot g_c$) for $V/D=5.0\cdot 10^{-3}=0.1T_K^0$, where
$T_K^0$ is the Kondo temperature in the absence of magnons ($g=0$). (b) An effective temperature
$T^{*}_{\chi}$ can be defined such that the $FDR_{\mathcal{\chi}}(\omega,T,V)$ of (a) collapses on
$\coth(\omega/2T)$, the equilibrium FDR of a bosonic field.
}
    \label{FDR_boson}
\end{figure*}

We now turn to the concept of effective temperature in the non-linear regime.
The notion of an effective temperature for extending the FDT
to non-equilibrium states was first introduced in the context
of steady states in 
chaotic systems~\cite{Hohenberg.89}, and was later used 
for non-stationary states in glassy 
systems~\cite{Cugliandolo.97}. For the ohmic spin-boson model, a variant of the BFKM, no meaningful
effective temperature can be defined~\cite{Mitra.05}. 
Here, we have an exact expression
for the FDR of the T-matrix in the steady state limit. 
It is easy to show that an effective temperature in the limit where $\omega\ll T$ (and $\alpha=1$)
can be obtained from Eq.~(\ref{fdr_g}):
\begin{eqnarray}
\tanh(\omega/T^{\star})&\equiv& FDR_{\mathcal{T}}(\omega,T,V)\Big|_{\omega\ll T} \nonumber \\
&\Rightarrow& T^{\star}_{\mathcal{T}}(T,V)=\frac{1}{2}(T+T \cosh(V/2T)).
\label{effT1}
\end{eqnarray}
This effective temperature restores the FDT for ${\mathcal{T}}(\omega,T,V)$ 
everywhere in the phase diagram irrespectively of whether
the system is critical or not and is ultimately connected to the boundary condition
that maintains the steady state, i.e.,
a time-independent particle flux through the magnetic SET. 
For no other correlator is the associated FDR 
completely determined by the steady state condition. In Ref.~\cite{Kirchner.09}, we argued that
generally, $FDR$s in the quantum critical regime of the magnetic SET displays both,  
$\omega/T$- and $V/T$-scaling. 
In order to address wether it is possible to introduce effective 
temperatures in the quantum critical regime 
for correlators other than ${\mathcal{T}}$,
we focus in the following
on the  dynamical spin response.
The dynamical spin susceptibility in the current-carrying steady state at arbitrary bias voltage 
$V$ has been explicitly
studied by an extension of the 
dynamical large-N of Ref.~\cite{Zhu.04} onto the Keldysh contour~\cite{Kirchner.09}.
Fig.~\ref{FDR_boson}(a) displays the  $FDR_{\mathcal{\chi}}(\omega,T,V)$ for 
$V=5\cdot 10^{-3}D=0.1 T_K^0$, with 
$T_K^0$ being the Kondo temperature in the absence of magnons ($g=0$).
The scaling regime extends up to energies of the order of $T_K^0$, so that bias voltages 
larger than $V=0.1 T_K^0$ might be affected by sub-leading contributions.
Three important observations underly the results displayed in Fig.~\ref{FDR_boson}(a):
(1) for $\omega>>T$, $FDR_{\mathcal{\chi}}(\omega,T,V)$ approaches the
value predicted by the FDT, ($\coth(\omega/2T)\rightarrow 1$),
(2) for $T>\approx 10^{-3}V$, the deviations from the linear response behavior are hardly 
discernible,
(3)  for $T<\approx 10^{-3}V$,it appears as if a simple scaling factor could collapse 
$FDR_{\mathcal{\chi}}(\omega,T,V)$ on the 
high-temperature curve, where (for $T\gg V$) the FDT applies.
(1)-(3) suggest, that an effective temperature of the form
\begin{equation}
T^{*}_{\chi}(T,V)=\frac{T}{\tanh\big(\alpha \frac{2 T}{V}\big)}
\end{equation}
could restore the FDT for all frequencies $\omega$.\\
As Fig.~\ref{FDR_boson}(b) demonstrates, this is indeed the case! The value of the scale factor
between $T$ and $V$ is $\alpha=10^{3}$, as suggested by observation (2). 
It is worth noting that the same $T^{*}_{\chi}(T,V)$ applies to other $V$ in the scaling regime
(with the same $\alpha$). The scaling factor $\alpha$ is determined by  some dimensionless
combination of the 'non-universal' parameters $D,g_{c},T_K^0$ and $\Lambda$.

While both effective temperatures are formally $\omega$-independent, $T^{*}_{\mathcal{T}}$
applies only in the limit $\omega\ll T$, and has a singular 
$T\rightarrow 0$-limit (with $V\neq 0$).
By contrast, $T^{*}_{\chi}\rightarrow V/(2\alpha)$ in the same limit.

In conclusion, we have shown that both, the (fermionic) T-matrix and the (bosonic) dynamical
spin susceptibility allow a description of their respective fluctuation-dissipation ratio
in terms of an effective temperature in the nonlinear, quantum critical regime (at least when
the probing frequency is much less than the temperature). 
The effective temperature $T^{*}$
is defined such that it restores the standard form of the FDT for
local (i.e. on the SET) fermionic
or bosonic variables  in the nonequilibrium steady-state limit.
The notion of temperature ($T$) in the present case is well defined as it
refers to the leads, which are kept at equal temperature.
Whether the concept of effective temperature in the quantum 
critical regime, applied here to the local T-matrix and spin-spin
correlator, can be extended to other correlators as well, will be
explored in a forthcoming publication.
\begin{acknowledgement}
This work has been supported in part by 
the NSF Grant No. DMR-0424125, the Robert A. Welch Foundation Grant
No. C-1411, and the W. M. Keck Foundation.
The authors gratefully acknowledge the hospitality 
of the Aspen Center 
for Physics where part of this work was completed.
\end{acknowledgement}

%
 \bibliographystyle{pss}
 \bibliography{nonequilib}

\providecommand{\WileyBibTextsc}{}
\let\textsc\WileyBibTextsc
\providecommand{\othercit}{}
\providecommand{\jr}[1]{#1}
\providecommand{\etal}{~et~al.}


\begin{thebibliography}{[10]}

\bibitem{Coleman.01}
 \textsc{P.~Coleman},  \textsc{C.~P\'{e}pin},  \textsc{Q.~Si},  and
  \textsc{R.~Ramazashvili},
 \jr{J.~Phys.~Cond.~Matt.} \textbf{13}, R723 (2001).


\bibitem{Si.01}
 \textsc{Q.~Si},  \textsc{S.~Rabello},  \textsc{K.~Ingersent},  and
  \textsc{J.~Smith},
 \jr{Nature} \textbf{413}, 804--808 (2001).


\bibitem{Loehneysen.07}
 \textsc{H.~v.~L\"{o}hneysen},  \textsc{A.~Rosch},  \textsc{M.~Vojta},  and
  \textsc{P.~W\"{o}lfle},
 \jr{Rev.~Mod.~Phys.} \textbf{79}, 1015 (2007).


\bibitem{Gegenwart.08}
 \textsc{P.~Gegenwart},  \textsc{Q.~Si},  and  \textsc{F.~Steglich},
 \jr{Nat.~Phys.} \textbf{4}, 186 (2008).


\bibitem{Kirchner.05a}
 \textsc{S.~Kirchner},  \textsc{L.~Zhu},  \textsc{Q.~Si},  and
  \textsc{D.~Natelson},
 \jr{Proc.~Natl.~Acad.~Sci.~USA} \textbf{102}, 18824--18829 (2005).


\bibitem{Kirchner.08b}
 \textsc{S.~Kirchner} and  \textsc{Q.~Si},
 \jr{Physica B} \textbf{403}, pp. 1189--1193 (2008).


\othercit
\bibitem{Kirchner.09}
 \textsc{S.~Kirchner} and  \textsc{Q.~Si},
Quantum criticality out of equilibrium: Steady state in a magnetic
  single-electron transistor,
arXiv:0805.3717, 2009.


\bibitem{Pasupathy.04}
 \textsc{A.\,N. Pasupathy},  \textsc{R.\,C. Bialczak},  \textsc{J.~Martinek},
  \textsc{J.\,E. Grose},  \textsc{L.\,A.\,K. Donev},  \textsc{P.\,L. McEuen},
  and  \textsc{D.\,C. Ralph},
 \jr{Science} \textbf{306}, 86--89 (2004).


\bibitem{Martinek.03}
 \textsc{J.~Martinek},  \textsc{M.~Sindel},  \textsc{L.~Borda},
  \textsc{J.~Barnas},  \textsc{J.~K\"{o}nig},  \textsc{G.~Sch\"on},  and
  \textsc{J.~von Delft},
 \jr{Phys.~Rev.~Lett.} \textbf{91}, 247202 (2003).


\bibitem{Choi.04}
 \textsc{M.~Choi},  \textsc{D.~Sanchez},  and  \textsc{R.~Lopez},
 \jr{Phys.~Rev.~Lett.} \textbf{92}, 056601 (2004).


\bibitem{Meir.92}
 \textsc{Y.~Meir} and  \textsc{N.~Wingreen},
 \jr{Phys.~Rev.~Lett.} \textbf{68}, 2512 (1992).


\bibitem{Hohenberg.89}
 \textsc{P.\,C. Hohenberg} and  \textsc{B.\,I. Shraiman},
 \jr{Physica D} \textbf{37}, 109--115 (1989).


\bibitem{Cugliandolo.97}
 \textsc{L.\,F. Cugliandolo},  \textsc{J.~Kurchan},  and
  \textsc{L.~Peliti},
 \jr{Phys.~Rev.~E} \textbf{55}, 3898 (1997).


\bibitem{Mitra.05}
 \textsc{A.~Mitra} and  \textsc{A.\,J. Millis},
 \jr{Phys.~Rev.~B.} \textbf{72}, 121102(R) (2005).


\bibitem{Zhu.04}
 \textsc{L.~Zhu},  \textsc{S.~Kirchner},  \textsc{Q.~Si},  and
  \textsc{A.~Georges},
 \jr{Phys.~Rev.~Lett.} \textbf{93}, 267201 (2004).


\end{thebibliography}
%

\end{document}